\documentclass[lettersize,journal]{IEEEtran}

\usepackage[utf8]{inputenc}
\usepackage[T1]{fontenc}
\usepackage{graphicx}
\usepackage[caption=false]{subfig}
\usepackage{xcolor}
\usepackage{amsmath,amssymb}
\usepackage{pdfpages}
\usepackage{acronym}
\usepackage{enumitem}
\usepackage{comment}
\usepackage{siunitx}
\usepackage{float}
\usepackage{longtable}
\usepackage{booktabs}
\usepackage{bm}
\usepackage{pgfgantt}
\usepackage{cite}
\usepackage[
    colorlinks=true,
    linkcolor=blue,
    citecolor=blue,
    urlcolor=blue
]{hyperref}
\usepackage{tabularx}
\usepackage[table]{xcolor}
\usepackage{ragged2e}
\usepackage[siunitx,american]{circuitikz}

\acrodef{RES}[RES]{Renewable Energy Source}
\acrodef{DER}[DER]{Distributed Energy Resource}
\acrodef{DG}[DG]{Distributed Generation}
\acrodef{PV}[PV]{Photovoltaic}
\acrodef{WT}[WT]{Wind Turbine}
\acrodef{EV}[EV]{Electric Vehicles}
\acrodef{CIR}[CIR]{Converter-Interfaced Resource}
\acrodef{CIG}[CIG]{Converter-Interfaced Generation}
\acrodef{SG}[SG]{Synchronous Generator}
\acrodef{SM}[SM]{Synchronous Machine}
\acrodef{PFC}[PFC]{Primary Frequency Control}
\acrodef{SFC}[SFC]{Secondary Frequency Control}
\acrodef{AGC}[AGC]{Automatic Generation Control}
\acrodef{LIPS}[LIPS]{Low-Inertia Power System}
\acrodef{CI-ESS}[CI-ESS]{Converter-Interfaced Energy Storage System}
\acrodef{ESS}[ESS]{Energy Storage System}
\acrodef{BESS}[BESS]{Battery Energy Storage System}
\acrodef{PHES}[PHES]{Pumped Hydroelectric Energy Storage}
\acrodef{FES}[FES]{Flywheel Energy Storage}
\acrodef{CAES}[CAES]{Compressed Air Energy Storage}
\acrodef{SMES}[SMES]{Superconducting Magnetic Energy Storage}
\acrodef{SCES}[SCES]{Supercapacitor Energy Storage}
\acrodef{P2G}[P2G]{Power-to-Gas}

\acrodef{water}[H\textsubscript{2}O]{Water}
\acrodef{hydrogen}[H\textsubscript{2}]{Hydrogen}
\acrodef{oxygen}[O\textsubscript{2}]{Oxygen}

\acrodef{HE}[HE]{Hydrogen Electrolyzer}
\acrodef{PEM}[PEM]{Proton Exchange Membrane}
\acrodef{FFR}[FFR]{Fast Frequency Response}
\acrodef{TS}[TS]{Transient Stability}

\acrodef{GFM}[GFM]{Grid-Forming}
\acrodef{GFL}[GFL]{Grid-Following}
\acrodef{GSP}[GSP]{Grid-Supporting}

\acrodef{PLL}[PLL]{Phase-Locked Loop}
\acrodef{PCC}[PCC]{Point of Common Coupling}

\acrodef{VSG}[VSG]{Virtual Synchronous Generator}
\acrodef{VSM}[VSM]{Virtual Synchronous Machine}
\acrodef{VOC}[VOC]{Virtual Oscillator}

\acrodef{RoCoP}[RoCoP]{Rate of Change of Power}
\acrodef{RoCoF}[RoCoF]{Rate of Change of Frequency}
\acrodef{SC}[SC]{Short-circuit}

\acrodef{LPF}[LPF]{Low-Pass Filter}
\acrodef{FRT}[FRT]{Fault Ride-Through}

\acrodef{PI}[PI]{Proportional-Integral}
\acrodef{PID}[PID]{Proportional-Integral-Derivative}

\acrodef{P}[P]{Active Power}
\acrodef{Q}[Q]{Reactive Power}

\acrodef{VSC}[VSC]{Voltage Source Converter}
\acrodef{WECS}[WECS]{Wind Energy Conversion System}

\acrodef{MPC}[MPC]{Model Predictive Control}

\acrodef{HV}[HV]{High-Voltage}
\acrodef{MV}[MV]{Medium-Voltage}
\acrodef{LV}[LV]{Low-Voltage}

\acrodef{GEM}[GEM]{Generalized ESS Model}
\acrodef{FACTS}[FACTS]{Flexible AC Transmission System}

\acrodef{SO}[SO]{System Operator}

\acrodef{EDL}[EDL]{Electrical Double Layer}
\acrodef{SoC}[SoC]{State of Charge}

\acrodef{AVR}[AVR]{Automatic Voltage Regulator}
\acrodef{HPR}[HPR]{High-Power Rectifiers}
\acrodef{CoI}[CoI]{Center of Inertia}
\acrodef{SVF}[SVF]{State-Variable Filter}
\acrodef{LS}[LS]{Least-Squares}
\begin{document}

\title{Modeling and Parametrization of Large-scale Hydrogen Electrolyzers for Transient and Frequency Stability Analysis}

\author{Mahmoud~K.~Abdelhamid,~\IEEEmembership{Graduate Student Member,~IEEE,}
        Lukas~Sigrist,~\IEEEmembership{Member,~IEEE,}
        and~Álvaro~Ortega,~\IEEEmembership{Member,~IEEE \vspace{-8mm}}%
\thanks{Mahmoud K. Abdelhamid is with the Institute for Research in Technology (IIT), ICAI School of Engineering, Comillas Pontifical University, 28015 Madrid, Spain, and also with the Electrical Engineering Department, Faculty of Engineering, Minia University, Minia 61111, Egypt (e-mails: mkhalifa@comillas.edu; mahmoud\_khalifa@mu.edu.eg).}%
\thanks{Lukas Sigrist and Álvaro Ortega are with the Institute for Research in Technology (IIT), ICAI School of Engineering, Comillas Pontifical University, 28015 Madrid, Spain (e-mail: \{lsigrist, aortega\}@comillas.edu).}%
\thanks{(Corresponding author: Mahmoud K. Abdelhamid.)}%
\thanks{The authors wish to thank Comunidad de Madrid, Spain, for the financial support to PREDFLEX project (TEC-2024/ECO-287) through the R\&D Activity Programme Tecnologías 2024.}%
}

\maketitle
\begin{abstract}
\acp{HE}, particularly \ac{PEM}-based systems, are flexible loads capable of providing grid ancillary services and enhancing the stability of renewable-dominated power systems due to their fast electrical response. However, the literature often lacks clear guidance on the structure, modeling, and parametrization of large-scale \acp{HE} for transient and frequency stability analyses. This paper presents a detailed configuration for utility-scale \ac{HE} plants, using the stack as the fundamental unit, supported by a realistic parametrization based on commercial components. Furthermore, this work extends the \ac{GEM}—a linearized state-space model previously utilized with other converter-interfaced energy storage systems—to \ac{PEM} \acp{HE} for the first time. To facilitate plant scalability, a robust method for scaling \ac{GEM} matrices across different capacities is proposed. Finally, a measurement-based methodology is developed to estimate the \ac{GEM} parameters and subsequently extract the equivalent electrical circuit parameters. Comprehensive simulations on the WSCC system under various disturbances validate the accuracy of the derived \ac{GEM}, the effectiveness of the scaling approach, and the robustness of the parameter identification methodology.
\end{abstract}
\acresetall  
\section{Introduction} 
\vspace{-2mm}
\subsection{Context}
Modern power systems are steadily being dominated by \acp{RES}, and hence affected by \ac{RES}'s inherent stochasticity, thus becoming more susceptible to active power imbalances and disturbances. \ac{P2G} solutions, such as \acp{HE}, couple electrical energy and gas systems and can provide needed load flexibility to mitigate the negative impact of \acp{RES} \cite{jain2023grid, TICHLER2016, WULF2018}. The flexibility stems from the ability to store \ac{hydrogen} in tanks and temporarily decouple its production and consumption \cite{pierluigi}.  

\acp{HE} can be classified into three main types: \ac{PEM}-based, alkaline-based, and solid oxide-based electrolyzers. Among these, \ac{PEM}-based electrolyzers present the fastest electrical response characteristics \cite{jain2023grid, van2024advanced} and full power adjustment on sub-second time scales has been demonstrated experimentally \cite{eichman2014novel}. This makes them suitable for efficiently providing grid ancillary services \cite{jain2023grid, pierluigi}. \ac{HE} was used in \cite{pierluigi, eichman2014novel, dozein2021fast2} as a flexible demand that can change its operating power and, thus, its \ac{hydrogen} production in response to frequency disturbances.

Two main modeling approaches can be distinguished for \ac{PEM}-based \ac{HE}. The first approach models the chemical reactions within cells and the detailed dynamics of key auxiliary equipment, such as compressors, pumps, and heat exchangers. Although this approach can yield a very accurate model, the resulting model is often too complex for power system stability studies and requires parameters that are often unavailable, especially for prospective studies \cite{han2015electrochemical, yigit2016mathematical, cooper2022framework}. The second approach is based on an equivalent electrical circuit model that reflects the electrolyzer's terminal DC voltage and current characteristics \cite{pierluigi}. This paper focuses on the modeling and parametrization of \ac{PEM}-based \ac{HE} using the second approach.\vspace*{-4mm}
\subsection{Related Literature}
\label{sub:intro_litrev}
The electrical models of \ac{PEM}-based \ac{HE} are widely used in power system studies and can be either static or dynamic. Static models are simplified models that represent only steady-state polarization curves and neglect the dynamics of the electrolyzer. On the other hand, dynamic models account for transient effects via capacitive elements that model the electrolyzer's dynamic behavior \cite{hernandez2020investigation, yodwong2020proton, ratib2024electrical}. In \cite{guilbert2019dynamic}, the performance of both static and dynamic models is compared with the experimental behavior of a real PEM \ac{HE}. The dynamic model achieved much higher accuracy than the static model, particularly during the transients.

Commonly, an aggregate electrical model is used in the literature to represent the dynamics of the \ac{HE} plant, consisting of the AC/DC and optionally DC/DC converters, and the \ac{HE}. Typical values of the parameters for this aggregate model can be found in several references, such as \cite{dozein2022virtual, dozein2021fast, dozein2021fast2}. \acp{HE} are built from stacks connected in strings (or sometimes called modules), with each stack comprising a number of cells. Practical commercial information is usually available at the stack and string levels \cite{Bosch_Hybrion_Stack, Siemens_Silyzer300}. However, the literature is often unclear about the fundamental unit actually modeled (cell, stack, or plant), and rarely specifies how many cells form a stack or how many stacks form the plant, complicating the mapping between \mbox{cell-,} stack-, and plant-level parameters. For instance, in \cite{ratib2025large}, a 1.044 MW \ac{HE} is formed from six \ac{HE} units; each unit is described only as \textquotedblleft one or more stacks\textquotedblright, but no information is given on how they are aggregated. In \cite{hossain2023dynamic}, a 400 W stack of \cite{guilbert2019dynamic} is treated as a single cell and scaled to a 1 MW stack of 2500 such units, without stating the number of series and parallel units. Recent works build plants from identical stacks or modules \cite{bernabeu2026modeling, huang2026aggregated, wijethunga2026frequency}. However, none of these works derive the stack arrangement from the voltage and current ratings of commercial AC/DC and DC/DC converters. This is a clear limitation for power system stability studies that require realistic and consistent plant parameters and configurations, such as those related to the optimal sizing and location of potential new ancillary service providers.

Moreover, while detailed dynamic models capture essential dynamics, their nonlinearity complicates parametrization. Several simplified models of other \acp{CI-ESS} have been introduced in the literature \cite{pal2000linear, fang2013design}, which could also be readily considered for the \ac{PEM}-based \acp{HE} considered in this work. However, these models represent storage dynamics solely by considering active and reactive power controllers, thereby neglecting the \acp{CI-ESS} dynamics. Consequently, as demonstrated in \cite{ortega2015generalized}, these simplified models are generally inaccurate and fail to capture the true system behavior during transients. To tackle the limitations of both highly complex and overly simplified models, the \ac{GEM} was proposed \cite{ortega2015generalized, bluebook}.

The \ac{GEM} simplifies the implementation and simulation of various \acp{CI-ESS} into a general and, more significantly for this work, linear form while maintaining accuracy. The \ac{GEM} can be derived for any \ac{CI-ESS} that is connected to the grid via an AC/DC converter and has a DC-link. While the \ac{GEM} has been derived for various \acp{CI-ESS} in the literature, it has not been applied to flexible loads like \acp{HE}. This is thus one of the main focuses of this work.

The main condition imposed in~\cite{ortega2015generalized} to derive the \ac{GEM} of a \ac{CI-ESS} is the availability of a \textit{detailed} dynamic model (and its parametrization) of said storage device. This is a clear limitation of the \ac{GEM}, since, and as discussed above, the availability of properly parametrized detailed dynamic models of  \ac{PEM}-based \acp{HE} is a rather strong assumption.
This paper addresses this gap and proposes an alternative derivation of the \ac{GEM} based on a reduced set of readily accessible measurements. Moreover, an algorithm to systematically and accurately scale the derived \ac{GEM} of \ac{PEM}-based \ac{HE} to different plant sizes is also developed and validated. \vspace*{-4mm}
\subsection{Contributions}
The paper makes the following contributions:
\begin{itemize}
    \item Modeling utility-scale \ac{PEM}-based \ac{HE} plants by using the \ac{HE} stack as a fundamental unit, with the stack arrangement derived from the voltage and current ratings of commercial AC/DC and DC/DC converters. Additionally, guidance on realistic and consistent parametrization of such models is provided by using practical commercial information.
    
    \item Deriving the \ac{GEM} for a \ac{PEM}-based \ac{HE} analytically. A scaling approach will also be proposed, enabling adaptation of the derived \ac{GEM} to any desired capacity. 
    
    \item Estimating the \ac{GEM} parameters from measurements while maintaining the physical interpretability of the parameters, which allows identifying the detailed model of the \ac{PEM}-based \ac{HE}.   
    \item Validating the \ac{GEM} of the \ac{PEM} \ac{HE} and the scaling approach under various operating conditions and disturbances.
\end{itemize}
The remainder of this study is structured as follows. Section \ref{sec:modleing_configuration} presents the detailed modeling of the \ac{HE} stack and the \ac{hydrogen} tank. Further, a realistic parametrization of such models is provided by making use of practical commercial information available. Section \ref{sec:GEM} presents the \ac{GEM} of a \ac{PEM} \ac{HE}. Additionally, the scaling approach is outlined to adapt the \ac{GEM} parameters to any capacity. Section \ref{sec:estimation} identifies the \ac{GEM} parameters using measurements and leverages these estimated matrices to extract the detailed model parameters. Section \ref{sec:case_studies} validates the \ac{GEM} and the estimated models by comparing their dynamic performance with that of the detailed model. Finally, conclusions are outlined in Section \ref{sec:conclusions}.
\section{System Modeling and Realistic Configuration}
\label{sec:modleing_configuration}
This section provides an overview of a \ac{PEM} \ac{HE} plant (Section~\ref{sub:model_ovw}), and introduces the detailed modeling of the \ac{PEM} \ac{HE} and the storage tank (Sections~\ref{sub:model_HE} and~\ref{sub:model_tank}) upon which the \ac{GEM} is based. Consistent and realistic parameters for the \ac{PEM} \ac{HE} plant and stacks are also given in Section~\ref{sec:Configuration}.\vspace*{-3mm} 
\subsection{Overview of a \ac{PEM} \ac{HE} Plant}
\label{sub:model_ovw}
\begin{figure*}[!htb]
    \centering
    \includegraphics[
        scale=0.46,
        trim={1cm 0cm 1cm 0cm},   
        clip
    ]{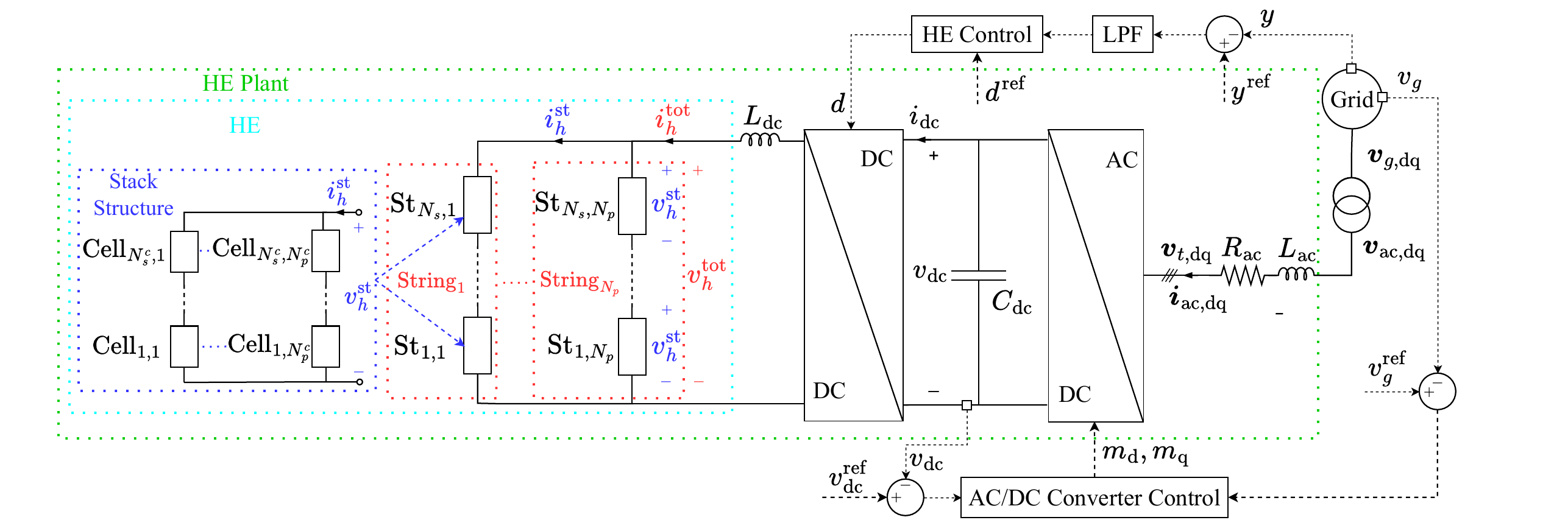}
    \caption{Complete topology of a grid-connected large-scale \ac{HE} plant.}
    \vspace{-2mm}
    \label{fig:realistic_configuration}
\end{figure*}

\ac{HE} plants consist of AC/DC and optionally DC/DC converters and the \ac{HE} (see Fig.~\ref{fig:realistic_configuration}). In case of AC/DC converters based on diode rectifiers, DC/DC converters are mandatory, whereas they are optional with IGBT rectifiers \cite{Secheron_WaterCooled_Converters}. The \ac{HE} is built of stacks connected in strings, where the stacks are made of cells connected in series and/or parallel \cite{agredano2024decentralized, ratib2024electrical, hernandez2020investigation, ursua2011hydrogen}. Although converters are represented in an aggregated manner for simplicity, in practice, multiple parallel converters are used \cite{Siemens_Silyzer300,  Siemens_Elyzerplant_100MW}. The AC/DC converter control has two channels: one for regulating the grid voltage ($v_g$) and the other for regulating the DC-link voltage ($v_\text{dc}$). The \ac{HE} control modulates the \ac{hydrogen} production rate according to a measured quantity of the network ($y$) by controlling the DC/DC converter. Both controllers typically use a standard \ac{PI} controller. The \ac{HE} plant is connected to the grid through a filter and a transformer.

The number of stacks in series in a string ($N_s$) is related to the rated DC voltage, whereas the number of parallel strings ($N_p$) is related to the rated DC current. Rated DC voltage and current also depend on the converters. Commercial \ac{PEM} \ac{HE} plants typically operate at \ac{LV} \cite{Siemens_Silyzer300, ABB_ACS880_Brochure, ABB_Rectifiers_Brochure, Siemens_Elyzerplant_100MW}. In general, DC voltage and current ratings of commercially available AC/DC rectifiers can reach 2 kV and hundreds of kA, respectively \cite{ABB_Rectifiers_Brochure}. Commercial DC/DC converters can withstand input voltages and currents up to 1.5 kV and 6 kA and supply \acp{HE} with up to 1.3 kV and 6 kA \cite{Ingeteam_H2_Brochure}. 
Congruently, the DC voltage and current ratings of large-scale \ac{HE} plants are in the range of 1 kV and tens to hundreds of kA.

The number of cells in series in a stack ($N_s^c$) is related to the stack voltage, which typically ranges around a few hundred volts~\cite{Bosch_Hybrion_Stack, agredano2024decentralized}.\vspace*{-3mm}

\subsection{Detailed Model of a \ac{PEM} \ac{HE}}
\label{sub:model_HE}
\begin{figure}[t]
    \centering
    \includegraphics[
        scale=0.6,
        trim={3cm 0.2cm 1cm 0cm},   
        clip
    ]{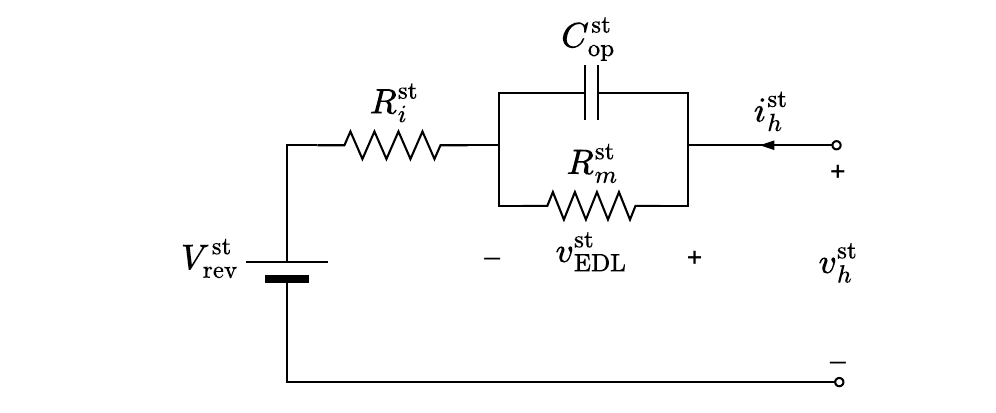}
    \caption{Dynamic model of a PEM \ac{HE} stack.}
    \vspace{-2mm}
    \label{fig:HE_model}
\end{figure}

If the responses of the individual stacks are similar, the total response of the \ac{HE} can be represented by an equivalent stack (see Fig.~\ref{fig:HE_model})~\cite{yodwong2020proton, hossain2023dynamic, huang2026aggregated}. The DC voltage source ($V_{\text{rev}}^\text{st}$) represents the reversible voltage, while $R_i^\text{st}$ represents the membrane resistance \cite{pierluigi, dozein2022virtual, yodwong2020proton}. It is worth mentioning that both $V_{\text{rev}}^\text{st}$ and $R_i^\text{st}$ can be considered either a function of the stack temperature and pressure \cite{pierluigi, dozein2021fast}, or a constant value \cite{hossain2023dynamic, ratib2025large} as it will be considered in this work.

The parallel connection of the over-potential capacitor ($C_{\text{op}}^\text{st}$) and the mass transport resistance ($R_m^\text{st}$) represents the \ac{EDL} phenomenon \cite{pierluigi, dozein2021fast, dozein2022virtual}.
The stack terminal voltage ($v_{h}^{\text{st}}$) can be expressed as follows:
\begin{equation}
\begin{split}
    v_{h}^{\text{st}}(t) = V_{\text{rev}}^\text{st} + i_{h}^{\text{st}}(t) R_i^\text{st} + v_{\text{EDL}}^\text{st}(t)
\end{split}
\label{eq:V_stack}
\end{equation}
where $v_{\text{EDL}}^\text{st}$ represents the voltage across the \ac{EDL} branch; and $i_{h}^{\text{st}}$ refers to the stack current, which can be defined as:
\begin{equation}
    i_h^{\text{st}} (t) = C_\text{op}^\text{st} \dot{v}_{\text{EDL}}^\text{st}(t) + \frac{v_{\text{EDL}}^\text{st}(t)}{R_m^\text{st}}
    \label{eq:I_stack}
\end{equation}

The hydrogen production rate for one stack ($n_{h}^{\text{st}}$), in mol/s, can be defined as \cite{pierluigi, castaneda2013sizing}:
\begin{equation}
    n_{h}^{\text{st}} (t)= \frac{N_s^c}{2F} \cdot \eta_F (t) \cdot i_{h}^{\text{st}} (t)
    \label{eq:n_h_stack}
\end{equation}
where $N_s^c$ is the number of series cells per stack \cite{castaneda2013sizing} and $F$ is the Faraday constant. The Faraday efficiency ($\eta_F$) is a non-linear function of the stack current and can be defined as \cite{dozein2022virtual, castaneda2013sizing, pierluigi}:
\begin{equation}
    \eta_F (t) = 0.965 \exp \left[ \frac{0.09}{i_{h}^{\text{st}}(t)} - \frac{75.5}{i_{h}^{\text{st}}(t)^2} \right]
    \label{eq:faraday}
\end{equation}

The production rate of the \ac{HE} plant is equal to the sum of the production rate of each stack. In the case of having $N_p$ parallel strings composed of $N_s$ series stacks each, as shown in Fig. \ref{fig:realistic_configuration}, then the total production rate of the \ac{HE} plant in mol/s ($n_{h}^{\text{tot}}$) can be computed as:
\begin{equation}
    n_{h}^{\text{tot}}(t)= N_s \cdot N_p \cdot n_{h}^{\text{st}}(t)
    \label{eq:n_h_total}
\end{equation}

Finally, the total terminal voltage and the total current of the plant, $v_h^{\text{tot}}$ and $i_h^{\text{tot}}$, are:
\begin{equation}
v_h^{\text{tot}}(t)=N_s \cdot v_h^{\text{st}}(t), \qquad
i_h^{\text{tot}}(t)=N_p \cdot i_h^{\text{st}}(t)
\label{eq:configuration1}
\end{equation}
\subsection{Storage Tank Model}
\label{sub:model_tank}
The hydrogen content in the tank is calculated by integrating the production rate from the total stacks ($n_{h}^{\text{tot}}$) and adding it to the tank's initial hydrogen content, if any \cite{pierluigi, castaneda2013sizing}:  
\begin{equation}
    m_h(t) = M_{h,0} + \int_{t_0}^{t} n_{h}^{\text{tot}}(t)\, dt - \int_{t_0}^{t} n_h'(t)\, dt
    \label{eq:M_h}
\end{equation}
where $m_h$ (\unit{mol}) and $M_{h,0}$ (\unit{mol}) are the hydrogen content in the tank and the tank's initial hydrogen content at $t =t_0$, respectively, while $n_h'(t)$ (\unit{mol/s}) is the hydrogen consumption rate. 
 
The \ac{SoC} can provide an indication for the hydrogen level inside the tank and can be defined as: 
\begin{equation}
    \mathrm{SoC}(t) = M_{\mathrm{molar}} \frac{m_h(t)}{M_{h}^{\text{max}}}
    \label{eq:SoC}
\end{equation}
where $M_{h}^{\text{max}}$ (\unit{kg}) represents the maximum value of the hydrogen content in the tank, while $M_{\mathrm{molar}}$ is the molar mass of hydrogen, which equals $0.002016$ \text{ kg/mol}.

The \ac{SoC} can also be related to the pressure inside the tank, $p_h$ (\unit{bar}), using the ideal gas equation \cite{castaneda2013sizing}:
\begin{equation}
\label{eq:pressure}
p_h(t)=10^{-5}\frac{\rho_hRT}{M_{\mathrm{molar}}}\,
\mathrm{SoC}(t)\,
\end{equation}
where $R$ (\unit{J.mol^{-1}.K^{-1}}) and $T$ (\unit{K}) are the ideal gas constant and the temperature inside the tank, respectively, while $\rho_h$ (\unit{kg/m^{3}}) represents the density of hydrogen gas at the tank's rated pressure and temperature.
\subsection{Realistic Parametrization of Utility-scale \ac{HE} Plants}
\label{sec:Configuration}
This section presents sets of parameters that are typical for real-world, large-scale \ac{PEM} \ac{HE} plants. Without loss of generality, a 100.5 MW \ac{HE} plant is considered as an example in this section. These parameters are consistent at the cell, stack, and plant levels.

The power-electronics interface of this realistic large-scale \ac{HE} plant is inspired by ABB ACS880-type AC/DC converters \cite{ABB_ACS880_Brochure} and Ingeteam H2 D-lyzer DC/DC converters \cite{Ingeteam_H2_Brochure}. To handle the overall power and current requirements of the 100.5 MW \ac{HE} plant, 17 pairs of these AC/DC and DC/DC converters can be used in parallel (100.5~kA/6~kA $\approx$ 17). 
The rating of the DC-link voltage ($v_\text{dc}$) is set to 1.2 kV, whereas the rated value of $v_h^{\text{tot}}$ is 1 kV. The aggregated DC/DC converter inductance, $L_\text{dc}$, and the DC-link capacitor, $C_\text{dc}$, are set to 0.1 mH \cite{dozein2022virtual, dozein2021fast} and 0.8 F \cite{zhu2012inertia}, respectively. 

The AC/DC converters are connected to the grid through a filter and a transformer. One or more transformer stages are needed to step down the AC grid voltage ($v_g$) to the required $v_\text{ac}$ as shown in Fig. \ref{fig:realistic_configuration}. For large-scale \ac{HE} plants, short-circuit impedance is around 0.15 pu. Typical values for filter parameters are $R_{ac} = 0.01$ pu and $L_{ac} = 0.2$ pu.

The parameters of the \ac{HE} stack adopted in this configuration are presented in Table \ref{tab:stack_parameters} \cite{agredano2024decentralized}, where $P_h^\text{st}, I_h^{\text{st}}$, and $V_h^{\text{st}}$ refer to the rated stack values of power, current, and voltage, respectively. We recall that the stack ratings were selected taking into account the characteristics of real industrial \ac{HE} stacks \cite{zhang2021wind, tuinema2020modelling}. Furthermore, Table \ref{tab:stack_parameters} introduces the cell parameters \cite{yodwong2020ac} utilized to evaluate the internal stack configuration: namely, the number of parallel branches within the stack ($N_p^c$), with each branch comprising $N_s^c$ series cells. 

From \eqref{eq:SoC}, it can be observed that determining the \ac{SoC} requires knowledge of $M_h^\text{max}$. Based on technical specifications from commercial \acp{HE} and utility-scale projects, the hydrogen production rate per MW of installed capacity is around $n_h^\text{max} = 18$ kg/h/MW \cite{Shell_100MW_HE, Bosch_Hybrion_Stack, Thyssenkrupp_20MW_HE, Siemens_Silyzer300, Siemens_Elyzerplant_100MW}, requiring for one hour of operational autonomy a storage capacity per installed capacity of around $M_h^\text{max} = 18$ kg/MW. If the tank's depth of discharge is limited, the storage capacity should increase \cite{ratib2025large}. 

The overall configuration of the 100.5 MW plant is summarized in Table \ref{tab:plant_configuration}. Note that the stack production rate ($n_h^\text{st,max}$) in Table \ref{tab:stack_parameters} is calculated using $I_h^{\text{st}}$ via \eqref{eq:n_h_stack}. This highlights the consistency of the proposed configuration, as multiplying $n_h^\text{st,max}$ by the total number of stacks precisely yields the aggregate plant production rate.
\begin{table}[!t]
    \centering
    \footnotesize                      
    \renewcommand{\arraystretch}{1.2}  
    \caption{Cell and Stack Model Parameters}
    \label{tab:stack_parameters}
    \begin{tabular}{lc | lc}
        \hline\hline
        \multicolumn{2}{c|}{\textbf{Cell}} & \multicolumn{2}{c}{\textbf{Stack}} \\
        Parameter & Value & Parameter & Value \\
        \hline
        Cell voltage  & 2 \unit{V} & $P_h^{\mathrm{st}}$ & 250 \unit{kW} \\
        Current density & 1 \unit{A/cm^2} & $I_{h}^{\mathrm{st}}$ & 1500 \unit{A} \\
        Cell area & 100 \unit{cm^2} & $V_{h}^{\mathrm{st}}$ & 166.67 \unit{V} \\
        & & $V_{\mathrm{rev}}^{\mathrm{st}}$ & 132 \unit{V} \\
        & & $R_i^{\mathrm{st}}$ & 5.79 \unit{m\ohm} \\
        & & $R_m^{\mathrm{st}}$ & 17.40 \unit{m\ohm} \\
        & & $C_{\mathrm{op}}^{\mathrm{st}}$ & 2.13 \unit{\farad} \\[0.3ex]
        \cline{3-4} 
        & & $N_s^c$ & 83 \\
        & & $N_p^c$ & 15 \\[0.5ex] 
        & & $n_h^\text{st,max}$ & 4.5 \unit{kg/h} \\[0.3ex]
        \hline\hline
    \end{tabular}
\end{table}  
\begin{table}[!t]
    \centering
    \footnotesize
    \renewcommand{\arraystretch}{1.1}
    \caption{Large-scale HE Plant Parameters (100.5 MW)}
    \label{tab:plant_configuration}
    \begin{tabular}{llc}
        \hline\hline
        \textbf{Parameter} & \textbf{Symbol} & \textbf{Value} \\
        \hline
        Number of series stacks per string & $N_s$ & 6 \\
        Number of parallel strings & $N_p$ & 67 \\
        Total (String) rated voltage & $V_h^{\text{tot}} = V_h^{\text{string}}$ & 1 \unit{kV} \\
        Total plant current & $I_h^{\text{tot}}$ & 100.5 \unit{kA} \\
        Total plant power & $P_h^{\text{tot}}$ & 100.5 \unit{MW} \\
        Total \ac{hydrogen} production rate & $n_h^\text{max}$ & 1809 \unit{kg/h} \\[0.3ex]
        \hline\hline\vspace*{-6mm}
    \end{tabular}
\end{table}
\section{\ac{GEM} for \ac{PEM} \ac{HE}}
\label{sec:GEM}
In this section, we first present an overview of the \ac{GEM} structure (Section~\ref{sub:gem_ovw}). Next, we analytically derive the \ac{GEM} from the detailed \ac{PEM} \ac{HE} (Section~\ref{sub:gem_he}). Since the \ac{GEM} parameters depend on the operating point, their adaptation to other operating points is then presented in Section~\ref{sub:gem_init}. Finally, we propose in Section~\ref{sub:gem_scaling} an approach that enables direct scaling of the \ac{GEM} matrices with respect to the capacity.
\vspace{-2mm}
\subsection{\ac{GEM} Structure}
\label{sub:gem_ovw}
The \ac{GEM} of a \ac{CI-ESS} can be formulated as the state-space linearization of its \textit{reduced} detailed model \cite{ortega2015generalized}:
\begin{equation}
\begin{aligned}
\bm{\mathrm{\tilde{T}}} \bm{\dot{x}}(t) &= \bm{\mathrm{\tilde{A}}} \bm{x(t)} + \bm{\mathrm{\tilde{b}}}_d d(t) + \bm{\mathrm{\tilde{b}}}_v v_\text{dc}(t) + \bm{\mathrm{\tilde{k}}}_x \\
i_\text{dc}(t) &= \bm{\mathrm{\tilde{c}}}^\text{T} \bm{x(t)} + \mathrm{\tilde{d}}_d d(t) + \mathrm{\tilde{d}}_v v_\text{dc}(t) + \mathrm{\tilde{k}}_i
\end{aligned}
\label{eq:GEM}
\end{equation}
where $\bm{x(t)}$ refers to the vector of two state variables of the detailed model, representing potential and flow, and for which the dynamics are preserved;  $d$ is the duty cycle determined by the \ac{HE} control, which is also considered as the \ac{GEM} input signal; and $i_\text{dc}(t)$ is the DC current flowing from the DC-link towards the storage, and its equation is considered the output equation of the \ac{GEM}.\vspace*{-3mm} 
\subsection{\ac{GEM} of the \ac{HE}}
\label{sub:gem_he}
The \ac{HE} together with the DC/DC converter is considered as the storage model, whereas 
the \ac{HE} control is the storage control in terms of \cite{ortega2015generalized}.  $d^\text{ref}$ is determined by the initial duty cycle, according to the initial power consumption of the electrolyzer (see Section \ref{sub:gem_init}). 

The buck and the boost operation modes of the bidirectional DC/DC converter can be expressed as follows \cite{bluebook}:
\begin{equation}
    \label{eq:DC/DC_converter}
    L_\text{dc} \dot{i}_h^{\text{tot}}(t)
    = -v_h^{\text{tot}}(t) + \left(1 - 2d(t)\right) v_\text{dc}(t)
\end{equation}
\begin{equation}
    i_\text{dc}(t) = \left(1 - 2d(t)\right) i_h^{\text{tot}} (t)
    \label{eq:DC/DC_converter2}
\end{equation}
Equations (\ref{eq:DC/DC_converter}) and (\ref{eq:DC/DC_converter2}) can be linearized around an operating point ($v_\text{dc,0}, d_0, i_{h,0}^{\text{tot}}$), and rewritten as:
\begin{equation}
\small
L_{\text{dc}} \dot{i}_h^{\text{tot}}(t)
= - v_h^{\text{tot}}(t)
- 2 v_{\text{dc},0}\, d(t)
+ (1 - 2 d_0)\, v_{\text{dc}}(t)
+ 2 d_0 v_{\text{dc},0}
\label{eq:linearized_dynamics}
\end{equation}
\begin{equation}
i_\text{dc}(t) = (1 - 2d_0)\, i_h^{\text{tot}}(t)
           - 2 i_{h,0}^{\text{tot}}\, d(t)
           + 2 d_0\, i_{h,0}^{\text{tot}}
\label{eq:linearized_dynamics2}
\end{equation}

Combining (\ref{eq:V_stack}) and (\ref{eq:I_stack}) and differentiating (\ref{eq:V_stack}) yields:
\begin{equation}
    \dot{v}_{\text{EDL}}^\text{st}(t) = \frac{i_h^{\text{st}}(t)}{C_{\text{op}}^\text{st}} - \frac{v_h^{\text{st}}(t) - V_{\text{rev}}^\text{st} - R_i^\text{st} i_h^{\text{st}}(t)}{C_{\text{op}}^\text{st} R_m^\text{st}}
\label{eq:edl_derivative}
\end{equation}
\begin{equation}
    \dot{v}_{\text{EDL}}^\text{st}(t) = \dot{v}_{h}^\text{st}(t) - R_i^\text{st} \dot{i}_h^\text{st}(t)
\label{eq:edl_derivative2}
\end{equation}

From (\ref{eq:edl_derivative}) and (\ref{eq:edl_derivative2}), the dynamic stack model can be derived and subsequently generalized for any capacity:
\begin{equation}
    C_{\text{op}}^{\text{tot}} (\dot{v}_h^{\text{tot}} - R_i^{\text{tot}} \dot{i}_h^{\text{tot}}) 
    = \frac{-v_h^{\text{tot}} + (R_i^{\text{tot}} + R_m^{\text{tot}}) i_h^{\text{tot}} + V_{\text{rev}}^{\text{tot}}}{R_m^{\text{tot}}}
    \label{eq:system_model}
\end{equation} 
where $v_h^{\text{tot}}$ and $i_h^{\text{tot}}$ can be calculated from (\ref{eq:configuration1}), while the stack model parameters can be scaled according to the number of series-parallel stacks in the following way:
\begin{equation}
\small
\begin{aligned}
    R_i^{\text{tot}} = \frac{N_s}{N_p} \cdot R_i^{\text{st}} & \qquad R_m^{\text{tot}} = \frac{N_s}{N_p} \cdot R_m^{\text{st}} \\
    V_{\text{rev}}^\text{tot} = N_s \cdot V_{\text{rev}}^\text{st} & \qquad C_{\text{op}}^\text{tot} = \frac{N_p}{N_s} \cdot C_{\text{op}}^\text{st}
\end{aligned}
\label{eq:GEM_parameters}
\end{equation}
\vspace{-2mm}

The detailed model for an \ac{HE} interfaced with a DC/DC converter can be thus expressed by (\ref{eq:DC/DC_converter}), (\ref{eq:DC/DC_converter2}), and (\ref{eq:system_model}). To obtain the corresponding \ac{GEM}, the system terminal voltage ($v_h^{\text{tot}}$) and current ($i_h^{\text{tot}}$) are selected as state variables (potential and flow, respectively). Using (\ref{eq:linearized_dynamics}), (\ref{eq:linearized_dynamics2}), and (\ref{eq:system_model}), it is straightforward to represent the model in the state-space form (\ref{eq:GEM}), and thus the following \ac{GEM} matrices can be derived:
\begin{equation}
\label{eq:GEM_matrices}
\begin{aligned}
\bm{x}(t) &= \begin{bmatrix} v_h^{\text{tot}}(t) \\ i_h^{\text{tot}}(t) \end{bmatrix}, \quad
\bm{X_{\text{init}}} = \begin{bmatrix} v_{h,0}^{\text{tot}} \\ i_{h,0}^{\text{tot}} \end{bmatrix}, \quad
\bm{\mathrm{\tilde{b}}}_{d} = \begin{bmatrix} 0 \\ -2 v_{\text{dc,0}} \end{bmatrix} \\
\bm{\mathrm{\tilde{b}}}_{v} &= \begin{bmatrix} 0 \\ 1 - 2 d_0 \end{bmatrix}, \quad
\bm{\mathrm{\tilde{T}}} = \begin{bmatrix} C_{\text{op}}^{\text{tot}} & -C_{\text{op}}^{\text{tot}} R_i^{\text{tot}} \\ 0 & L_{\text{dc}} \end{bmatrix} \\
\bm{\mathrm{\tilde{A}}} &= \begin{bmatrix} -\dfrac{1}{R_m^{\text{tot}}} & 1 + \dfrac{R_i^{\text{tot}}}{R_m^{\text{tot}}} \\ -1 & 0 \end{bmatrix}, \quad
\mathrm{\tilde{d}}_{d} = -2 i_{h,0}^{\text{tot}}, \quad
\mathrm{\tilde{k}}_{i} = 2 d_0 i_{h,0}^{\text{tot}} \\
\bm{\mathrm{\tilde{k}}}_{x} &= \begin{bmatrix} \dfrac{V_{\text{rev}}^{\text{tot}}}{R_m^{\text{tot}}} \\ 2 d_0 v_{\text{dc,0}} \end{bmatrix}, \quad
\bm{\mathrm{\tilde{c}}}^{\text{T}} = \begin{bmatrix} 0 & 1 - 2 d_0 \end{bmatrix}, \quad
\mathrm{\tilde{d}}_{v} = 0
\end{aligned}
\end{equation}
where the vector $\bm{X_\text{init}}$ includes the initial values of the \ac{GEM} state variables. 

One of the main benefits of the \ac{GEM}, namely the model order reduction, does not directly apply in this case, as the number of states coincides in both models. However, the benefits of the linear structure of the equivalent \ac{GEM} become apparent, particularly when designing advanced control algorithms for these systems, or when running large sets of simulations, such as in Monte Carlo analysis. 

The \ac{SoC} can be expressed in terms of the state variables of the \ac{GEM}. Without loss of generality, $n_h'$ is assumed to be zero. Therefore, (\ref{eq:M_h}) can be reformulated as:
\begin{equation}
m_{h}(t) = M_{h,0} + k_m N_s N_p \int_{t_0}^{t} \eta_{F} (t) \cdot i_{h}^{\text{st}}(t) \, dt
\label{eq:M_h_int}
\end{equation}
where $k_m = \frac{N_s^c}{2F}$. 
By rewriting (\ref{eq:M_h_int}) in derivative form and linearizing, we get:
\begin{equation}
\small
\dot{m}_{h}(t) = k_m N_s N_p \left[ (\eta_{F,0} + i_{h,0}^{\text{st}} \eta'_{F,0}) i_{h}^{\text{st}}(t) - \left( i_{h,0}^{\text{st}} \right)^2 \eta'_{F,0}\right]
\label{eq:M_h_lin}
\end{equation}
where 
\begin{equation}
\small
    \begin{aligned}
\eta_{F,0} &= 0.965 \exp \left[ \frac{0.09}{i_{h,0}^{\text{st}}} - \frac{75.5}{(i_{h,0}^{\text{st}})^2} \right], \\
\eta'_{F,0} &= \eta_{F,0} \left[ \frac{-0.09}{(i_{h,0}^{\text{st}})^2} + \frac{151}{(i_{h,0}^{\text{st}})^3} \right]
    \end{aligned}
    \label{eq:faraday_lin}
\end{equation} 

Consequently, by integrating (\ref{eq:M_h_lin}), expressing it in terms of the state variables via (\ref{eq:GEM_matrices}) and (\ref{eq:configuration1}), and substituting the result into (\ref{eq:SoC}), the \ac{SoC} can be directly reformulated as:
\begin{equation}
\mathrm{SoC}(t) = \mathrm{SoC}_0 +\sum_{i=1}^{2} \bm{k_i} \int_{t_0}^{t} \left[\bm{x}_i(t) -\bm{x}_{c,i} \right]
\label{eq:SoC_GEM2}
\end{equation}
where $\mathrm{SoC}_0= M_{\mathrm{molar}} M_{h,0}/M_h^{\max}$, $\bm{x}_i$ represents the \ac{GEM} state variables; while $\bm{k}_i$ and $\bm{x}_{c,i}$ can be defined as:
\begingroup
\small
\begin{equation}
\label{eq:SoC_matrices}
\begin{aligned}
\bm{k}_i &=
\begin{bmatrix}
0 \\
\frac{M_{\mathrm{molar}} k_m N_s (\eta_{F,0} + i_{h,0}^{\text{st}} \eta'_{F,0})}
       {M_h^{\max}}
\end{bmatrix}, \quad
\bm{x}_{c,i} =
\begin{bmatrix}
0 \\
\frac{N_p \left(i_{h,0}^{\text{st}}\right)^2 \eta'_{F,0}}
     {\eta_{F,0} + i_{h,0}^{\text{st}} \eta'_{F,0}}\vspace*{-3mm}
\end{bmatrix}
\end{aligned}
\end{equation}
\endgroup
\vspace{-2mm}
\subsection{Dependency of \ac{GEM} on the Operating Point}
\label{sub:gem_init}
From (\ref{eq:GEM_matrices}), it can be observed that the \ac{GEM} matrices strongly depend on the operating point ($i_{h,0}^{\text{tot}}$ and $d_0$), which can be estimated from the initial AC power consumption of the \ac{HE} plant, $P_0$ (which is not much affected by the power flow solution).  

The power balance for the given AC/DC converter in Fig.~\ref{fig:realistic_configuration} can be formulated as \cite{bluebook}:
\begin{equation}
p_\text{ac} (t) = p_\text{dc} (t) + C_\text{dc} v_\text{dc}(t) \dot{v}_\text{dc}(t) + p_\text{loss}
    \label{eq:VSC_balance}
\end{equation}
where $p_\text{ac}$ and $p_\text{loss}$ represent the AC active power at \ac{PCC} and ohmic power losses caused by $R_\text{ac}$, respectively, while $p_\text{dc}$ refers to the DC power, which can be defined as: $p_\text{dc} (t) = v_\text{dc} (t) \cdot i_\text{dc} (t)$. 

The initial DC current ($i_\text{dc,0}$) can be obtained from (\ref{eq:VSC_balance}) by setting $p_\text{ac} = P_0$ as follows:
\begin{equation}
i_\text{dc,0} =\frac{P_0 - p_\text{loss}}{v_\text{dc,0}} 
    \label{eq:idc_0}
\end{equation}
where $v_\text{dc,0}$ is the initial value of the DC-link voltage, equal to the reference value, i.e., $v_\text{dc,0} = v_\text{dc}^\text{ref}$. Another expression for $i_\text{dc,0}$ can be found from (\ref{eq:DC/DC_converter2}):
\begin{equation}
    i_\text{dc,0} = \left(1 - 2d_0\right) i_{h,0}^{\text{tot}}
    \label{eq:idc_0_}
\end{equation}
The initial total voltage ($v_{h,0}^\text{tot}$) and the initial total current ($i_{h,0}^\text{tot}$) can be expressed by using (\ref{eq:DC/DC_converter}) and (\ref{eq:system_model}):
\begin{equation}
v_{h,0}^{\text{tot}} = \left(1 - 2d_0\right) v_\text{dc,0}
    \label{eq:v_h_0}
\end{equation} 
\begin{equation}
\small
i_{h,0}^{\text{tot}} = \frac{\left(1 - 2d_0\right) v_\text{dc,0} - V_{\text{rev}}^{\text{tot}}}{R_i^{\text{tot}} + R_m^{\text{tot}}} 
    \label{eq:i_h_0}
\end{equation} 
From (\ref{eq:idc_0_}) and (\ref{eq:i_h_0}), we can obtain:
\begin{equation}
i_\text{dc,0} = \left(1 - 2d_0\right) \left[ \frac{\left(1 - 2d_0\right) v_\text{dc,0} - V_{\text{rev}}^{\text{tot}}}{R_i^{\text{tot}} + R_m^{\text{tot}}} \right] 
    \label{eq:i_dc_0_final}
\end{equation}

Firstly, $i_\text{dc,0}$ is calculated using (\ref{eq:idc_0}), then (\ref{eq:i_dc_0_final}) can be solved for $d_0$. Since (\ref{eq:i_dc_0_final}) is quadratic,  the valid solution must satisfy the condition $0<d_0<0.5$, implying power consumption by the \ac{HE}. After obtaining $d_0$, it is straightforward to find $i_{h,0}^{\text{tot}}$ using (\ref{eq:i_h_0}). Then, $v_{h,0}^{\text{tot}}$ can be found using (\ref{eq:v_h_0}), while the initial stack voltage ($v_{h,0}^{\text{st}}$) and initial stack current ($i_{h,0}^{\text{st}}$) can be obtained using the same analogy of (\ref{eq:configuration1}). Finally, $\eta_{F,0}$ and $\eta'_{F,0}$ can be calculated using (\ref{eq:faraday_lin}).
\subsection{Scaling of the \ac{GEM} According to the Capacity} 
\label{sub:gem_scaling}
In Section~\ref{sub:intro_litrev}, we discussed how obtaining a realistic and reliable set of parameters for different \ac{HE} plant capacities is arguably a challenging process, if possible at all. In this regard, we present here a scaling method to estimate the \ac{GEM} matrices for any desired capacity ($P_h^\text{new}$) by using \ac{GEM} matrices of a given capacity ($P_h^\text{giv}$), for which accurate parameters are assumed to be known. Hence, the full set of vectors and matrices in~\eqref{eq:GEM_matrices} is considered as given, and will be denoted with the superscript ``giv'' in the remainder of this Section. Essentially, the scaling method consists of modifying some elements of the \ac{GEM} matrices by appropriate scaling factors that depend on the given ($N_s^\text{giv}$ and $N_p^\text{giv}$) and the new number ($N_s^\text{new}$ and $N_p^\text{new}$) of series stacks and parallel strings. 

The number of new required stacks ($N_\text{stacks}^\text{new}$) is found based on the new capacity ($P_{h}^\text{new}$) and the stack's rated power ($P_h^\text{st}$): $N_\text{stacks}^\text{new} = P_{h}^\text{new}/P_h^\text{st}$. After that, the number of series stacks per string ($N_s^\text{new}$) is selected based on the nominal voltage of the AC/DC converter, then the number of parallel strings is selected such that: $N_p^\text{new} = N_\text{stacks}^\text{new}/N_s^\text{new}$. Afterward, two scaling factors (one for the voltage and the other for the current) can be defined as follows:
\begin{equation}
K_s = \frac{N_s^{\text{new}}}{N_s^{\text{giv}}}, \quad
K_p = \frac{N_p^{\text{new}}}{N_p^{\text{giv}}}
\label{eq:scaling_factor2}
\end{equation}
Note that when the AC/DC converter is similar to the given one in terms of power capacity and voltage ratings, the value of $N_s^{\text{new}}$ should be as close as possible to $N_s^\text{giv}$ to keep $V_h^{\text{tot}}$ close to its reference value. To achieve this, the following conditions should be satisfied: $N_p^\text{new} = N_\text{stacks}^\text{new}/N_s^{\text{giv}}$ where $N_p^\text{new} \geq 1$, then $N_s^\text{new} = N_\text{stacks}^\text{new}/N_p^{\text{new}}$.

The given initial duty cycle, $d_0^\text{giv}$, can be determined using the second element of $\bm{\mathrm{\tilde{b}}}_{v}^\text{giv}$ as follows: \mbox{$d_0^{\text{giv}} = 0.5 \left( 1 - \mathrm{b}_{v,2}^{\text{giv}} \right)$}. From (\ref{eq:v_h_0}), one can obtain \mbox{$v_{h,0}^{\text{giv}} = Y^{\text{giv}} \cdot v_\text{dc,0}$} and \mbox{$v_{h,0}^{\text{new}}\!=\!Y^{\text{new}}\cdot v_\text{dc,0}$} where $Y^{\text{giv}}\!=\!1-2d_0^{\text{giv}}$ and \mbox{$Y^{\text{new}}\!=\!1-2d_0^{\text{new}}$}. The DC-link voltage is maintained constant by the AC/DC converter controller as shown in Fig. \ref{fig:realistic_configuration}, while $v_{h,0}^{\text{new}}$ scales linearly with $N_s^{\text{new}}$, therefore, $v_{h,0}^{\text{new}} = K_s \cdot v_{h,0}^{\text{giv}}$, which leads to $Y^{\text{new}} = K_s \cdot Y^{\text{giv}}$. Finally, the new initial duty cycle can be defined as: $d_0^{\text{new}} = 0.5(1-Y^{\text{new}})$. 

Adding stacks in series scales the total voltage and equivalent impedance, while adding strings in parallel scales the total current and equivalent admittance. Therefore, $R_i, R_m$ and $L_{\mathrm{dc}}$ are scaled by the factor $\frac{K_s}{K_p}$, while $C_{\mathrm{op}}$ and $\frac{1}{R_m}$ are scaled by the factor $\frac{K_p}{K_s}$. 

Regarding the \ac{SoC} matrices, it is evident from (\ref{eq:SoC_matrices}) that $\bm{k}_{i}$ primarily depends on $N_s$ and on the tank capacity $M_h^{\max}$, while $\bm{x}_{c,i}$ mainly depends on $N_p$; therefore, they are scaled using the factors $K_s/K_t$ and $K_p$, respectively, where $K_t = M_h^{\max,\text{new}}/M_h^{\max,\text{giv}}$. If the tank is not resized, $K_t = 1$ and $\bm{k}_{i}$ is scaled by $K_s$ only.

Consequently, the scaled \ac{GEM} matrices can be expressed as in (\ref{eq:GEM_scaling2}), where the scaling method provided is valid for scaling the \ac{GEM} matrices to any desired capacity relative to a given capacity. 
\begin{equation}
\small
\label{eq:GEM_scaling2}
\begin{aligned}
\bm{\mathrm{\tilde{T}}}^\text{new} &=
\begin{bmatrix}
T_{11}^\text{giv}\cdot K_p/K_s & T_{12}^\text{giv}\\
0 & T_{22}^\text{giv}\cdot K_s/K_p
\end{bmatrix}, \quad
\bm{\mathrm{\tilde{b}}}_{d}^\text{new}=\bm{\mathrm{\tilde{b}}}_{d}^\text{giv},\\
\bm{\mathrm{\tilde{A}}}^\text{new} &=
\begin{bmatrix}
A_{11}^\text{giv}\cdot K_p/K_s & A_{12}^\text{giv}\\
-1 & 0
\end{bmatrix}, \quad
\bm{\mathrm{\tilde{b}}}_{v}^\text{new}=
\begin{bmatrix}0\\1-2d_0^\text{new}\end{bmatrix},\\
\bm{\mathrm{\tilde{k}}}_{x}^\text{new} &=
\begin{bmatrix}
k_{x,1}^\text{giv}\cdot K_p\\
k_{x,2}^\text{giv}\cdot d_0^\text{new}/d_0^\text{giv}
\end{bmatrix}, \quad
\bm{x}_{c,i}^\text{new}=
\begin{bmatrix}0\\x_{c,2}^\text{giv}\cdot K_p\end{bmatrix},\\
\bm{\mathrm{\tilde{c}}}^{\mathrm{T,new}} &=
\begin{bmatrix}0&1-2d_0^\text{new}\end{bmatrix}, \quad
\bm{k}_{i}^\text{new}=\bm{k}_{i}^\text{giv}\cdot K_s/K_t,\\
\mathrm{\tilde{d}}_{d}^\text{new} &= \mathrm{\tilde{d}}_{d}^\text{giv}\cdot K_p,\quad
\mathrm{\tilde{d}}_{v}^\text{new}=0,\quad
\mathrm{\tilde{k}}_{i}^\text{new}
=\mathrm{\tilde{k}}_{i}^\text{giv}\cdot K_p\cdot d_0^\text{new}/d_0^\text{giv}.
\end{aligned}
\end{equation}

The process can be simplified when the desired capacity is an integral multiple of the given capacity and an AC/DC converter with the same voltage rating is assumed. Therefore, in this scenario, scaling is performed exclusively through the parallel strings based on the target capacity, while keeping the number of series stacks constant. Consequently, we have $N_s^{\text{new}} = N_s^\text{giv}$, which means that $V_h^{\text{tot}}$ remains constant. Additionally, \(v_{\text{dc},0}\) is maintained at a constant value, as previously mentioned. From (\ref{eq:v_h_0}), we can conclude that $d_0^\text{new} = d_0^\text{giv}$.

Finally, the scaling factors in (\ref{eq:scaling_factor2}) can be updated as follows:
\begin{equation}
K_s = \frac{N_s^{\text{new}}}{N_s^{\text{giv}}} = 1, \quad
K_p = \frac{N_p^{\text{new}}}{N_p^{\text{giv}}} = \frac{P_h^\text{new}}{P_h^\text{giv}}
\label{eq:scaling_factor1}
\end{equation}
It is clear from (\ref{eq:scaling_factor1}) that knowledge of the reference configuration ($N_s^\text{giv}$ and $N_p^\text{giv}$) is not necessary in that case, as the scaling can be done directly using the power ratios. By applying (\ref{eq:scaling_factor1}) and substituting $d_0^\text{new} = d_0^\text{giv}$ in (\ref{eq:GEM_scaling2}), we can get the final and simple scaling form for that case.
\section{Model Parameters Identification Using Measurements}
\label{sec:estimation}
It has already been discussed that one of the limitations of the \ac{GEM} is the required availability of a complete, detailed dynamic model of the storage system under study. This can prove to be a rather strong assumption, and difficult to achieve in practice. In this regard, Section~\ref{sub:estimation_gem} introduces a methodology for estimating the \ac{GEM} matrices using a limited set of readily available measurements. With regard to the \ac{PEM} \acp{HE} studied in this work, only the voltages and currents of both sides of the DC/DC converter are needed without requiring measurements at the stack level. Equivalent measurements are needed if other converter-interfaced technologies (batteries, supercapacitors, flywheels, etc.) are considered. The proposed methodology also shows another interesting property, as it allows the development of a direct and straightforward method for extracting the detailed model parameters from the estimated \ac{GEM} matrices, which is presented in Section~\ref{sub:estimation_detailed}.
\subsection{\ac{GEM} Matrices Estimation}
\label{sub:estimation_gem}
For the developments presented in this Section, it is assumed that the following measurements are available: $v_\text{dc}(t)$, $i_\text{dc}(t)$, $v_h^{\text{tot}}(t)$, and $i_h^{\text{tot}}(t)$. Then $d^\text{est}(t)$ can be directly estimated using these measurements from (\ref{eq:DC/DC_converter2}) as follows: 
\begin{equation}
d^\text{est}(t)=0.5\left[1-i_{\mathrm{dc}}(t)/i_h^{\mathrm{tot}}(t)\right]
\label{eq:duty_estimated}
\end{equation}
Further, it is straightforward to get the initial values $v_\text{dc,0}$ and $d_0$, and therefore use~\eqref{eq:GEM_matrices} to estimate $\text{b}_{d,2}^\text{est} = -2 v_\text{dc,0}$ and $\text{b}_{v,2}^\text{est}= 1 - 2 d_0^\text{est}$. 

The first step towards the \ac{GEM} matrices estimation involves rewriting the equation of the state variables in (\ref{eq:GEM}) in incremental form:
\begin{equation}
    \bm{\mathrm{\tilde{T}}}\, \Delta \bm{\dot{x}}(t)
    =
    \bm{\mathrm{\tilde{A}}}\, \Delta \bm{x}(t)
    + \bm{\mathrm{\tilde{b}}}_d\, \Delta d(t)
    + \bm{\mathrm{\tilde{b}}}_v\, \Delta v_{\mathrm{dc}}(t)
    \label{eq:incremental}
\end{equation}
where $\Delta x_1(t)$ and $\Delta x_2(t)$ are the increments of the signals $v_h^{\text{tot}}(t)$, and $i_h^{\text{tot}}(t)$. By knowing $\bm{\mathrm{\tilde{b}}}_d$ and $ \bm{\mathrm{\tilde{b}}}_v$, and since $T_{21} = A_{22} = b_{d1} = b_{v1} = 0$ and $A_{21} = -1$ (see~\eqref{eq:GEM_matrices} and~\eqref{eq:GEM_scaling2}), (\ref{eq:incremental}) can be written as $\bm{\Phi}(t)\bm{\theta} = \bm{R}(t)$ \cite{Garnier2008}, with
\begin{align}
\bm{\Phi}(t) &=
\begin{bmatrix}
\Delta \dot{x}_1(t) & \Delta \dot{x}_2(t) & 0 & -\Delta x_1(t) & -\Delta x_2(t) \\
0 & 0 & \Delta \dot{x}_2(t) & 0 & 0
\end{bmatrix}, \nonumber\\
\bm{\theta} &=
\begin{bmatrix}
T_{11}^\text{est} & T_{12}^\text{est} & T_{22}^\text{est} &
A_{11}^\text{est} & A_{12}^\text{est}
\end{bmatrix}^{\mathrm T}, \nonumber\\
\bm{R}(t) &=
\begin{bmatrix}
0 \\
-\Delta x_1(t)+b_{d2}\Delta d(t)+b_{v2}\Delta v_{\mathrm{dc}}(t)
\end{bmatrix}.
\label{eq:regression_terms}
\end{align}

$\Delta \dot{x}_1(t)$ and $\Delta \dot{x}_2(t)$ can be approximated from the sampled discrete measurements \cite{Garnier2008}. 

The regression \eqref{eq:regression_terms} is then solved in \ac{LS} sense for $\bm{\theta}$. The remaining terms ($\bm{\mathrm{\tilde{b}}}_d^\text{est}$,
$\bm{\mathrm{\tilde{b}}}_v^\text{est}$, $\bm{\mathrm{\tilde{c}}}^\text{T,est}$,
$\mathrm{\tilde{d}}_d^\text{est}$, $\mathrm{\tilde{d}}_v^\text{est}$ and
$\mathrm{\tilde{k}}_i^\text{est}$) are found directly from the operating point
and \eqref{eq:GEM_matrices}, and
$\bm{\mathrm{\tilde{k}}}_x^\text{est}$ from \eqref{eq:GEM} in steady state.

Note that the estimated elements ($T_{11}^\text{est}$, $T_{12}^\text{est}$, $A_{11}^\text{est}$, and $A_{12}^\text{est}$) might, however, be scaled by an arbitrary and unknown factor since the first equation of the regression model is homogeneous. To overcome this problem, an analytical correction factor ($\alpha$) is defined that multiplies the first-row elements:
\begin{equation}
\label{eq:correction_factor}
\alpha =
\left(A_{12}^{\mathrm{est}} -
A_{11}^{\mathrm{est}}T_{12}^{\mathrm{est}}/T_{11}^{\mathrm{est}}\right)^{-1}
\end{equation}

Indeed, substituting the \ac{GEM} matrix elements from (\ref{eq:GEM_matrices}) into (\ref{eq:correction_factor}) shows that $\alpha$ should equal 1.

With noisy measurements, the derivative approximation amplifies the noise, and the estimate degrades severely (see Table~\ref{tab:noise_comparison}). In this case, specific techniques, such as the \ac{SVF}, exist and can be applied to avoid such problems \cite{Garnier2008}. 

Finally, although the estimation focuses on the \ac{GEM}, the same approach can be used to estimate the \ac{HE} control parameters.
\subsection{Estimation of the Detailed Model Parameters}
\label{sub:estimation_detailed}
Referring to (\ref{eq:GEM_matrices}), the detailed model parameters in Fig. \ref{fig:HE_model} can be estimated from the estimated \ac{GEM} matrices, as follows:
\begin{equation}
\label{eq:estimated_detailed_model}
\begin{aligned}
C_{\mathrm{op}}^{\mathrm{tot,est}} &= T_{11}^{\mathrm{est}}, \quad
L_{\mathrm{dc}}^{\mathrm{est}} = T_{22}^{\mathrm{est}}, \quad
R_{\mathrm{m}}^{\mathrm{tot,est}} = -\frac{1}{A_{11}^{\mathrm{est}}}, \\
V_{\mathrm{rev}}^{\mathrm{est}} &= -\frac{\tilde{k}_{x,1}^{\mathrm{est}}}{A_{11}^{\mathrm{est}}}, \quad
R_{\mathrm{i}}^{\mathrm{tot,est}} = -\frac{T_{12}^{\mathrm{est}}}{T_{11}^{\mathrm{est}}}
= \frac{1-A_{12}^{\mathrm{est}}}{A_{11}^{\mathrm{est}}}.
\end{aligned}
\end{equation}
\section{Case Studies}
\label{sec:case_studies}
This section presents and discusses a comprehensive case study that validates:
\begin{enumerate}
    \item the \ac{GEM} derived from the detailed dynamic model of the \ac{HE} by comparing the dynamic performance of both models in different scenarios (Section~\ref{sub:case_gem});
    \item the methodology proposed to estimate the \ac{GEM} from current and voltage measurements. Its robustness is also tested under a variety of disturbances (Section~\ref{sub:case_estim}); and
    \item the methodology to scale the \ac{HE} to any capacity  (Section~\ref{sub:case_scaling}). 
\end{enumerate}
 
\begin{figure}[!t]
    \centering
    \includegraphics[
        scale=0.3,
        trim={0cm 0.2cm 0cm 0cm},   
        clip
    ]{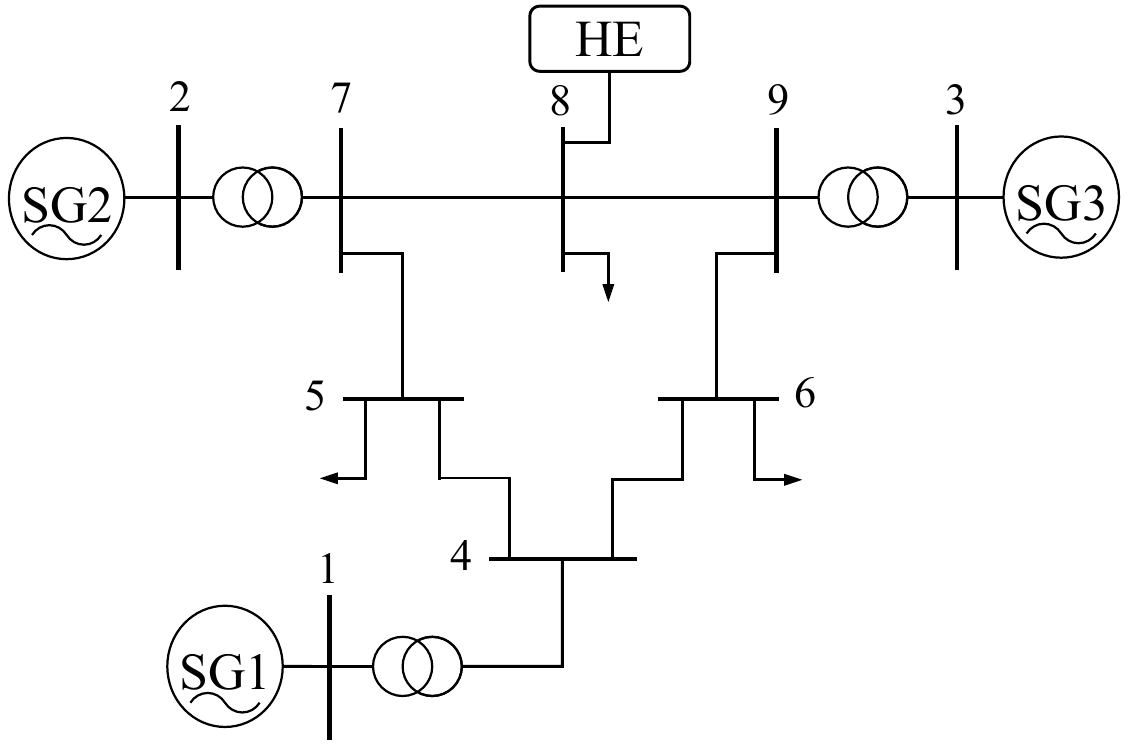}
    \caption{IEEE WSCC 9-bus system with \ac{HE} connected to bus 8.}
    \label{fig:WSCC}
\end{figure}

Although the developed \ac{HE} model can be integrated with any general power grid, all simulations in this work are conducted using the WSCC 9-bus system shown in Fig. \ref{fig:WSCC}. A 100.5 MW \ac{HE}, described in Table \ref{tab:plant_configuration}, is connected to bus 8 via converters and a transformer that steps down the grid voltage to 600 V, while maintaining the total load at bus 8 similar to the original static load of the WSCC system. The \ac{HE} control in Fig. \ref{fig:realistic_configuration} regulates the frequency of the \ac{CoI}. Unless otherwise indicated, the \ac{HE} operates at an initial power of 0.5 pu, while $\mathrm{SoC}_0$ is set to 0.2 pu. The control system parameters are provided in Table \ref{tab:control_parameters}.
\begin{table}[!bt]
    \centering
    \footnotesize                      
    \renewcommand{\arraystretch}{1.2}  
    \caption{Control System Parameters}
    \label{tab:control_parameters}
    \begin{tabular}{lc}
        \hline\hline
        \textbf{Parameter} & \textbf{Value} \\
        \hline
        \multicolumn{2}{c}{\textbf{AC/DC Converter Control}} \\
        Outer DC voltage controller & $0.12 + 24/s$ \\
        Outer AC voltage controller & $5 + 1000/s$ \\
        Inner current control loops & $0.3 + 10/s$ \\[1ex] 
        
        \multicolumn{2}{c}{\textbf{HE Control}} \\
        Controller gains & $3 + 10^{-4}/s$ \\
        Low-pass filter (LPF) & $1/(1+s)$ \\
        \hline\hline
    \end{tabular}
\end{table}
\subsection{\ac{GEM} Validation}
\label{sub:case_gem}
This subsection is dedicated to validating the derived \ac{GEM} of the \ac{HE} versus its detailed model. 
\subsubsection{Contingency}
\begin{figure}[!t]
    \centering
    
    \includegraphics[width=0.48\columnwidth]{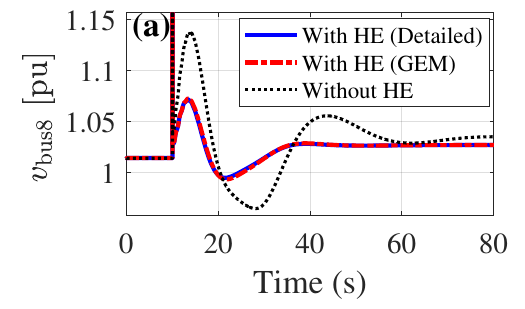}\hfil
    \includegraphics[width=0.48\columnwidth]{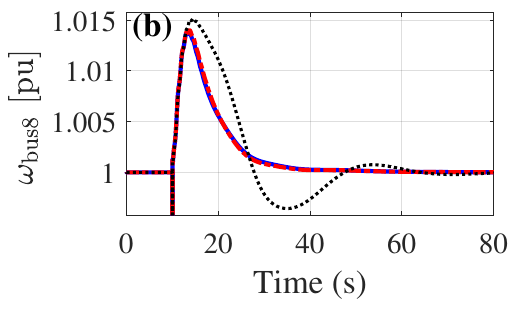}
    
    \vspace{-0.45cm}
    
    \includegraphics[width=0.48\columnwidth]{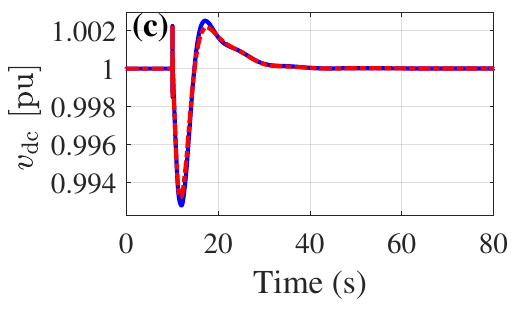}\hfil
    \includegraphics[width=0.48\columnwidth]{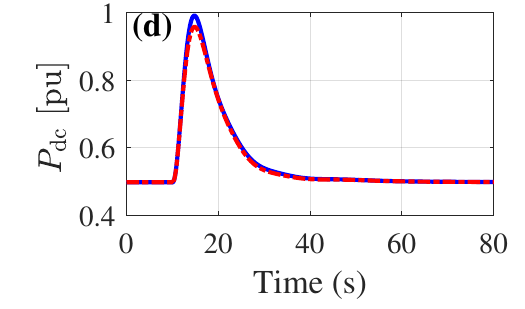} 
    
    \vspace{-0.45cm}
    
    \includegraphics[width=0.48\columnwidth]{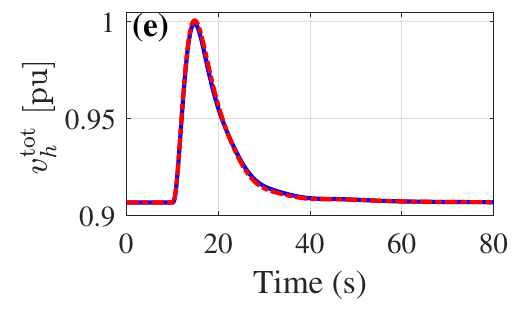}\hfil
    \includegraphics[width=0.48\columnwidth]{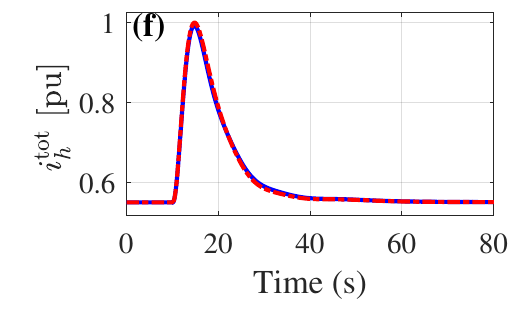}
    
    \vspace{-0.2cm}
    \caption{Dynamic response comparison of the detailed model and the \ac{GEM} under a loss of load at bus 5.}
    \label{fig:GEM_validation_cont1_G2}
\end{figure}

The load at bus~5, rated at 125~MW and 50~MVAR (representing approximately 39\% of the total system generation), is disconnected at $t = 10$~s to evaluate the dynamic performance of the \ac{GEM}. The results are shown in Fig. \ref{fig:GEM_validation_cont1_G2}, which overall reveal that the response of the \ac{GEM} is very close to that of the detailed model. 

Figures~\ref{fig:GEM_validation_cont1_G2}(a) and~\ref{fig:GEM_validation_cont1_G2}(b) show and compare the voltage and the frequency dynamics at bus~8 ($v_\text{bus8}$ and $\omega_\text{bus8}$) for the two models. Both models show remarkably similar responses. Further, the case without \ac{HE} is included to show that adding the \ac{HE} improves voltage and frequency dynamics (smaller overshoot and reduced settling time).

Figures~\ref{fig:GEM_validation_cont1_G2}(c) and~\ref{fig:GEM_validation_cont1_G2}(d) show and compare the voltage at the DC-link and the DC power. It can be observed that both models exhibit very similar responses, with a very small mismatch at the peaks during the transient due to the linearization underlying the \ac{GEM}. This will be further discussed later on. Note that the power consumption increases due to the \ac{HE} control acting on the frequency deviation.  

Finally, $v_h^{\text{tot}}$ and $i_h^{\text{tot}}$ are shown and compared in Figs.~\ref{fig:GEM_validation_cont1_G2}(e) and~\ref{fig:GEM_validation_cont1_G2}(f), respectively. These results again reveal that the \ac{GEM} response is very well aligned with the response of the detailed model. 
This strong alignment occurs because the dynamic stack model (\ref{eq:system_model}) is linear in both cases. Furthermore, because both models share near-identical grid frequency responses, their controllers generate identical duty cycles ($d$); combined with the DC-link voltage remaining very close to its nominal value ($v_{\text{dc}} \approx v_{\text{dc},0}$), the error introduced by linearizing the DC/DC converter equations is negligible.
\subsubsection{Dynamic Setpoint Change} 
\label{sec:dynamic_set_point}
In this case study, the synchronous generator at bus~2 is replaced by a wind turbine subjected to stochastic wind perturbations following a Weibull distribution. The \ac{GEM} and detailed model are tested under normal dynamic operation. At $t = 50$~s, the electrolyzer's power setpoint is gradually ramped down from $0.5$~pu to $0.2$~pu (by adjusting the DC/DC converter's duty cycle). The main objective of this test is to examine the validity of the \ac{GEM} at another operating point, different from the one at which the matrices are derived. 

\begin{figure}[!t]
    \centering
    
    \includegraphics[width=0.48\columnwidth]{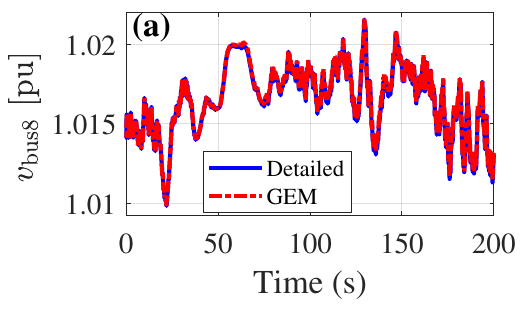}\hfil
    \includegraphics[width=0.48\columnwidth]{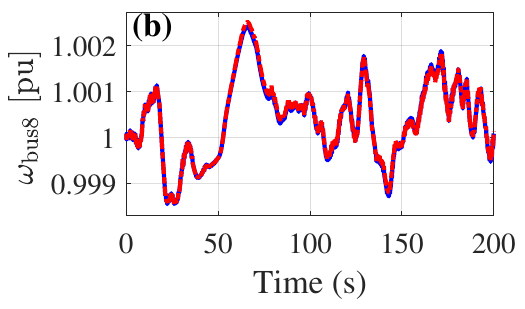}
    
    \vspace{-0.45cm}
    
    \includegraphics[width=0.48\columnwidth]{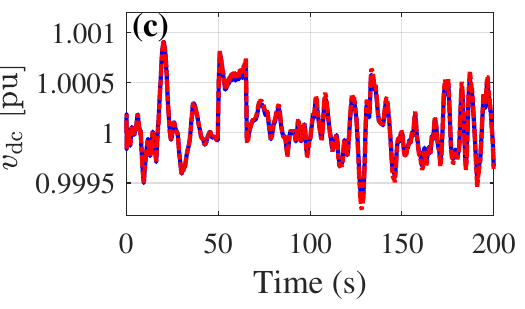}\hfil
    \includegraphics[width=0.48\columnwidth]{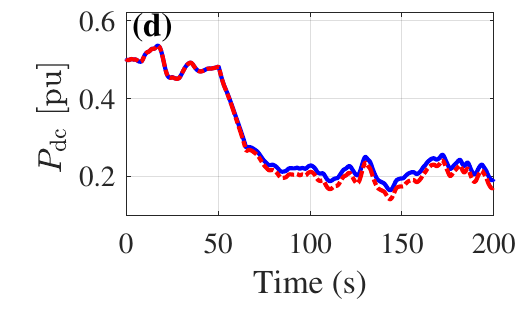} 

    \vspace{-0.45cm}

    \includegraphics[width=0.48\columnwidth]{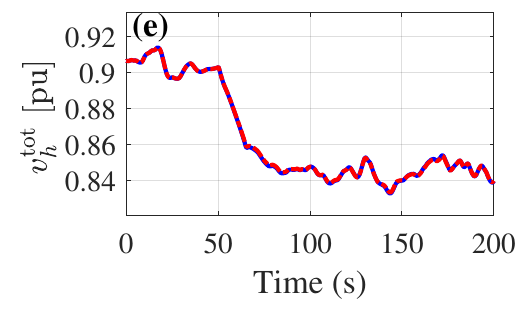}\hfil
    \includegraphics[width=0.48\columnwidth]{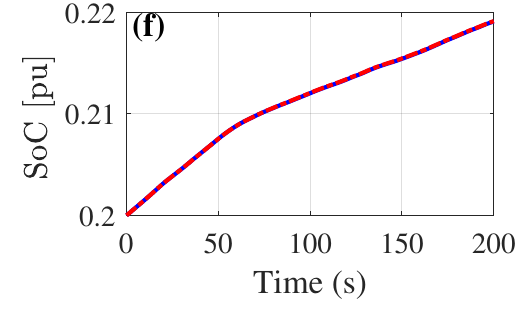}
    
    \vspace{-0.2cm}
    \caption{Dynamic performance comparison of both models during a dynamic setpoint change considering stochastic wind perturbations.}
    \label{fig:GEM_Validation_Performance_Cont0_WT}
\end{figure}

Figures~\ref{fig:GEM_Validation_Performance_Cont0_WT}(a) and~\ref{fig:GEM_Validation_Performance_Cont0_WT}(b) present the dynamics of $v_\text{bus8}$ and $\omega_\text{bus8}$, respectively, in which the two models are nearly aligned throughout the stochastic wind variations and even after applying the setpoint change. This demonstrates that shifting the operating point introduces negligible deviations in the \ac{GEM}'s representation of the AC signals at the \ac{PCC}.

Figure~\ref{fig:GEM_Validation_Performance_Cont0_WT}(c) illustrates the variation of $v_{\text{dc}}$, which is regulated by the AC/DC converter controller. The controller successfully maintains $v_{\text{dc}}$ close to its reference value. Figure~\ref{fig:GEM_Validation_Performance_Cont0_WT}(d) shows and compares $p_{\text{dc}}$, where the impact of the linearization around the initial set point of 0.5 pu can be seen after the set point change, although with minimal differences (\textasciitilde  0.02 pu).

Finally, Figs.~\ref{fig:GEM_Validation_Performance_Cont0_WT}(e) and ~\ref{fig:GEM_Validation_Performance_Cont0_WT}(f) show and compare $v_h^{\text{tot}}$ and the \ac{SoC}, where the dynamic responses of both models are perfectly aligned. Therefore, it can be concluded that the derived \ac{GEM} for \ac{PEM} \ac{HE} can successfully reproduce the dynamic response of the detailed model. Although the \ac{GEM} is a linearized model derived at a specific operating point, it shows a minor mismatch relative to the detailed model even under significant setpoint changes.
\subsection{Estimation Validation}
\label{sub:case_estim}
This subsection validates the proposed estimation approach. The model parameters are estimated by using synthetic measurements generated via a time-domain simulation of the adopted power system shown in Fig. \ref{fig:WSCC}, including the detailed \ac{HE} model. For the estimation, a loss-of-load at bus 5 has been simulated. 

Table~\ref{tab:noise_comparison} presents a systematic comparison of parameter identification under ideal and noisy conditions. Using noiseless measurements (LS Ideal), the estimation is highly accurate, yielding matrices nearly identical to their references. However, when measurements are corrupted with a peak-to-peak noise of 2--3\% of their rated values, the standard \ac{LS} estimation fails  (LS Noisy), as discussed in Section \ref{sub:estimation_gem}. Applying an \ac{SVF} effectively mitigates this issue (LS+SVF Noisy), restoring the estimation accuracy. Although $T_{22}$ (and consequently $L_{\text{dc}}$) exhibits a relative error of approximately 24\% in both the ideal and SVF cases, the absolute deviation remains negligible (${\sim}0.0024$~pu). This is corroborated by the identified models' ability to accurately reproduce the detailed model's dynamic response, as discussed below.
\begin{table}[!t]
\renewcommand{\arraystretch}{1.15}
\setlength{\tabcolsep}{2.5pt}
\caption{Identified Parameters Under Ideal and Noisy Measurements.}
\label{tab:noise_comparison}
\centering
\scriptsize
\begin{tabular}{l r r r r}
\hline\hline
 & & \multicolumn{1}{c}{LS} & \multicolumn{1}{c}{LS} & \multicolumn{1}{c}{LS+SVF}\\
\multicolumn{1}{c}{Param.} & \multicolumn{1}{c}{Ref.} &
\multicolumn{1}{c}{Ideal} & \multicolumn{1}{c}{Noisy} &
\multicolumn{1}{c}{Noisy}\\
\hline
$T_{11}$        & $0.2367$  & $0.2368\,(0.02\%)$   & $6.7\mathrm{e}{-5}$   & $0.2277\,(-3.8\%)$\\
$T_{12}$        & $-0.0123$ & $-0.0123\,(0.36\%)$  & $-1.5\mathrm{e}{-10}$ & $-0.0112\,(-9.1\%)$\\
$T_{22}$        & $0.0100$  & $0.0124\,(24.5\%)$   & $5.9\mathrm{e}{-5}$   & $0.0125\,(24.6\%)$\\
$A_{11}$        & $-6.3857$ & $-6.3869\,(0.02\%)$  & $-4.9250$             & $-6.2658\,(-1.9\%)$\\
$A_{12}$        & $1.3328$  & $1.3330\,(0.02\%)$   & $1.0000$              & $1.3077\,(-1.9\%)$\\
\hline
$C_{op}^{tot}$  & $0.2367$  & $0.2368\,(0.02\%)$   & $6.7\mathrm{e}{-5}$   & $0.2277\,(-3.8\%)$\\
$R_{i}^{tot}$   & $0.0520$  & $0.0521\,(0.34\%)$   & $2.3\mathrm{e}{-6}$   & $0.0491\,(-5.5\%)$\\
$R_{m}^{tot}$   & $0.1566$  & $0.1566\,(-0.02\%)$  & $0.2030$              & $0.1596\,(1.9\%)$\\
$V_{rev}^{tot}$ & $0.7921$  & $0.7920\,(-0.01\%)$  & $0.7969$              & $0.7920\,(-0.01\%)$\\
$L_{dc}$        & $0.0100$  & $0.0124\,(24.5\%)$   & $5.9\mathrm{e}{-5}$   & $0.0125\,(24.6\%)$\\
\hline\hline
\end{tabular}
\end{table}

\begin{figure}[!t]
    \centering

    \includegraphics[width=0.48\columnwidth]{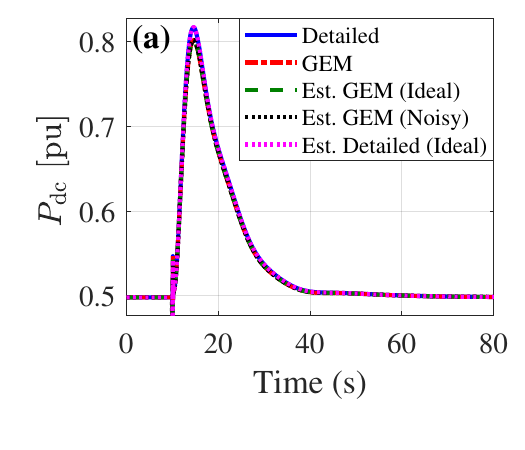}\hfil 
    \includegraphics[width=0.48\columnwidth]{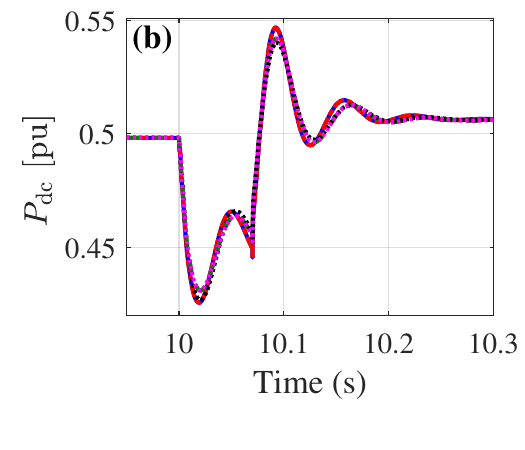}

    \vspace{-0.78cm} 

    \includegraphics[width=0.48\columnwidth]{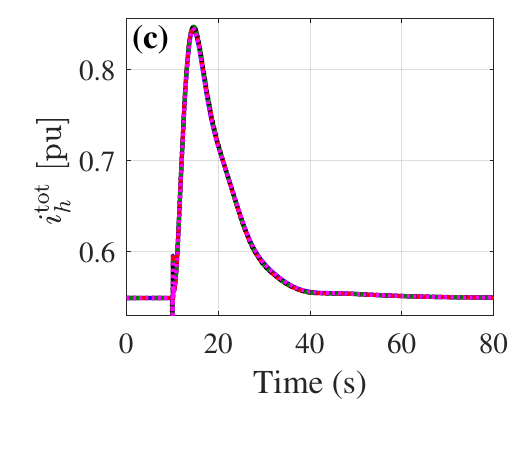}\hfil
    \includegraphics[width=0.48\columnwidth]{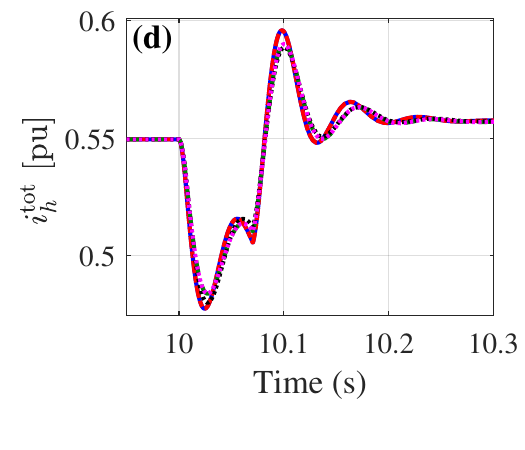}
    
    \vspace{-0.5cm}
    \caption{Validating the estimated models during a cascading disturbance (\ac{SC} at bus 6 followed by loss of load at the same bus).}
    \label{fig:Estimation_Robustness_Cont7_G2}
\end{figure}

To evaluate the robustness of the estimated parameters, the estimated models are validated and compared with the actual detailed model and the analytically derived \ac{GEM} for various faults and contingencies different from the one used for the estimation. As a representative case, a severe cascading disturbance is presented. As shown in Fig. \ref{fig:Estimation_Robustness_Cont7_G2}, a \ac{SC} fault is applied at bus 6 at $t = 10$~s. The fault is cleared after $70$~ms by isolating the line between buses 4 and 6, which is immediately followed by a loss-of-load at bus 6 at $t = 10.07$~s.

In Fig.~\ref{fig:Estimation_Robustness_Cont7_G2}, the identified models from noiseless measurements are labeled ``Ideal'', while the model identified from corrupted measurements via the \ac{SVF} is labeled as ``Noisy''.
Figure~\ref{fig:Estimation_Robustness_Cont7_G2}(a) compares the dynamic response of $p_{\text{dc}}$, whereas Fig.~\ref{fig:Estimation_Robustness_Cont7_G2}(b) provides a zoom during the \ac{SC}. The responses of the estimated and initial models match. Compared to the detailed model, the maximum identical-mismatch error is $1.5\%$ in the analytical \ac{GEM} and both estimated \acp{GEM}, whereas the estimated detailed model shows an error of about $0.6\%$.

Figures~\ref{fig:Estimation_Robustness_Cont7_G2}(c) and~\ref{fig:Estimation_Robustness_Cont7_G2}(d) illustrate the overall dynamic response and the detailed transient view of $i_h^{\text{tot}}$, respectively. Both the analytical \ac{GEM} and the estimated models tightly track the detailed model, with a maximum mismatch of $0.45\%$ for the analytical \ac{GEM} and around $1\%$ across all estimated models. 
\subsection{\ac{GEM} Matrices Scaling Validation}
\label{sub:case_scaling}
To validate the proposed \ac{GEM} scaling approach, a 60~MW \ac{HE} is implemented using 40 parallel strings ($N_p = 40$) of the 1.5~MW string ($N_s = 6$) described in Section \ref{sec:Configuration}. The scaling formulation in (\ref{eq:GEM_scaling2}) is applied to scale the reference \ac{GEM} matrices of the 100.5~MW capacity down to the 60~MW level. The dynamic performance of this mathematically scaled \ac{GEM} is then evaluated against both the detailed 60~MW model and its analytically derived \ac{GEM}.

The dynamic setpoint case study from Section \ref{sec:dynamic_set_point} is replicated to evaluate the scaled 60~MW \ac{GEM} under identical stochastic wind perturbations and a ramped power reduction from 0.5~pu to 0.2~pu at $t = 50$~s. The dynamic performance of the scaled \ac{GEM} is then compared against both the analytically derived $60\text{ MW}$ \ac{GEM} and its detailed non-linear model. 

\begin{figure}[!t]
    \centering
    
    \includegraphics[width=0.48\columnwidth]{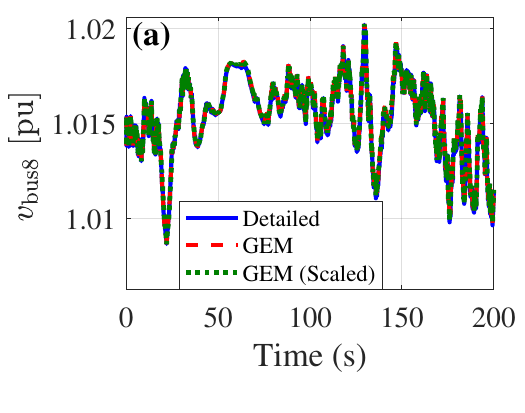}\hfil
    \includegraphics[width=0.48\columnwidth]{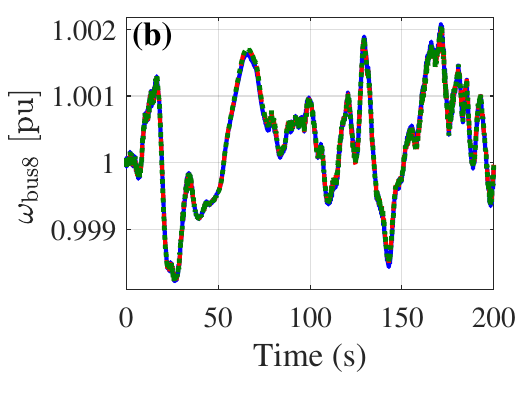}
        
    \vspace{-0.7cm}
    
    \includegraphics[width=0.48\columnwidth]{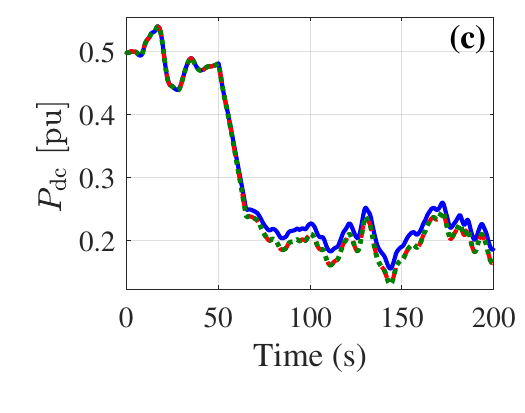}\hfil
    \includegraphics[width=0.48\columnwidth]{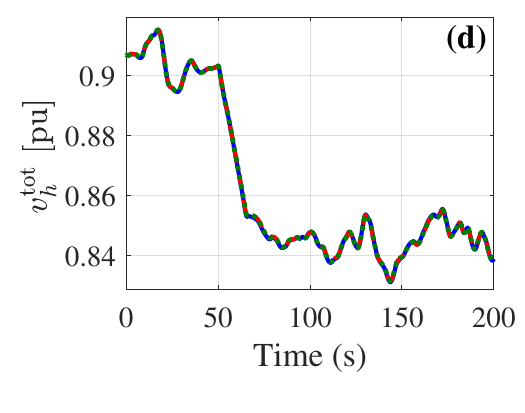}
    
    \vspace{-0.4cm}
    \caption{Dynamic performance comparison of the 60~MW \ac{HE} during a dynamic setpoint change considering stochastic wind perturbations.}
    \label{fig:GEM_Scaling_Performance_Cont0_WT}
\end{figure}

The results are shown in Fig. \ref{fig:GEM_Scaling_Performance_Cont0_WT}. As observed, $v_{\text{bus8}}$, $\omega_{\text{bus8}}$ and $v_h^{\text{tot}}$ maintain perfect alignment across the three models; this tight tracking is also maintained in $v_{\text{dc}}$ and $i_h^{\text{tot}}$. Conversely, a mismatch error of approximately $2.5\%$ is identically exhibited by both the scaled and analytical \acp{GEM} with respect to the detailed model in $i_{\text{ac,d}}$, $i_{\text{dc}}$, and $p_{\text{dc}}$ during the setpoint change. Because the dynamic response of the scaled \ac{GEM} completely overlaps with the analytically derived \ac{GEM}, and both exhibit the same minor deviations observed with respect to the detailed model during large setpoint changes, the proposed scaling methodology is successfully validated.
\section{Conclusions}
\label{sec:conclusions}
This paper has addressed the modeling and parametrization of utility-scale \ac{PEM} \acp{HE} to facilitate transient and frequency stability analysis in modern power systems. To achieve this, a practical large-scale \ac{HE} plant configuration was established with a realistic and consistent model parametrization applicable to any capacity, and the \ac{GEM}—a linear state-space representation—was derived for \acp{HE}. Furthermore, a generalized scaling methodology for \ac{GEM} matrices and a measurement-based model parameter estimation approach were proposed. A primary benefit of the proposed estimation approach is that it relies on a few easy-to-obtain measurements, enabling direct extraction of detailed equivalent electrical circuit parameters and overcoming the challenge of limited manufacturer data.

The validation results demonstrate that both the derived and scaled \ac{GEM} models accurately replicate the detailed nonlinear model dynamics during normal operation and grid transients. Furthermore, the robustness of the estimation methodology is validated under a severe cascading disturbance, in which the estimated \ac{GEM} and detailed model behave closely to their analytical and detailed counterparts, respectively.

Future work will focus on extending this measurement-based identification approach to other \acp{CI-ESS} and determining the required measurements for those technologies. Additionally, future efforts will explore implementing more sophisticated estimation techniques and utilizing real-world operational measurements, if available.
\bibliographystyle{IEEEtran} 
\bibliography{Refs} 

\end{document}